\documentclass{article}
\usepackage{ijcai26}

\usepackage{times}
\usepackage{soul}
\usepackage{xurl}
\usepackage[hidelinks]{hyperref}
\usepackage[utf8]{inputenc}
\usepackage[small]{caption}
\usepackage{graphicx}
\usepackage{amsmath}
\usepackage{amsthm}
\usepackage{booktabs}
\usepackage{algorithm}
\usepackage{algorithmic}
\usepackage[switch]{lineno}
\usepackage{amssymb}
\usepackage{tipa}

\usepackage{multirow} 
\usepackage{array}    

\usepackage{colortbl}
\usepackage{xcolor}
\usepackage[hang,flushmargin]{footmisc}

\definecolor{bestgray}{gray}{0.90}
\definecolor{secondgray}{gray}{0.95}

\title{Profiling the Voice: Speaker-Specific Phoneme Fingerprinting for Speech Deepfake Detection}

 \author{
 Jun Xue$^{1,2}$\and
 Tong Zhang$^{1,2}$\and
  Zhuolin Yi$^{1,2}$\and
 Yihuan Huang$^{1,2}$\and
 Yi Chai$^{1,2}$\and
 Yiyang Zhang$^{1,2}$\And
 Yanzhen Ren$^{1,2}$\thanks{Corresponding author.}
 \affiliations
  $^1$Key Laboratory of Aerospace Information Security and Trusted Computing, Ministry of Education\\
 $^2$School of Cyber Science and Engineering, Wuhan University\\
 \emails
 \{junxue, tongzhang, renyz\}@whu.edu.com,
 }

\begin{document}

\maketitle


\begin{abstract}
The rapid advancement of generative AI has made audio deepfakes increasingly indistinguishable from authentic human vocals, posing significant threats to persons-of-interest (POI) such as public figures. Current detection systems primarily rely on generic, black-box models that fail to capture speaker-specific idiosyncratic traits and lack interpretability. In this paper, we propose Phoneme-based Voice Profiling (PVP), a novel personalized defense framework. By shifting the detection paradigm from macro-utterance analysis to micro-phonetic modeling, PVP captures the unique acoustic distributions underlying a POI’s habitual articulatory patterns. Specifically, our framework models speaker-specific phonetic realizations using lightweight Gaussian Mixture Models (GMMs) estimated solely from bona fide reference speech. This design enables data-efficient profiling and robust generalization to previously unseen spoofing attacks without requiring heavy spoof-specific training. Furthermore, we introduce the first large-scale Chinese POI deepfake dataset to benchmark speaker-specific detection. Experimental results demonstrate that PVP significantly outperforms state-of-the-art generic detectors in POI spoofing scenarios, achieving substantial EER reductions while providing fine-grained, phoneme-level interpretability for forensic analysis. Code and data are available at: \url{https://github.com/JunXue-tech/PVP}
\end{abstract}

\section{Introduction}

With the rapid development of artificial intelligence generated content (AIGC), speech synthesis and voice conversion technologies have achieved unprecedented realism, making speech deepfakes increasingly accessible and convincing. As a result, speech deepfake detection has become a critical research topic, especially in high-stakes scenarios such as judicial forensics, governmental communications, and public statements by celebrities or other well-known figures. In these contexts, a single forged utterance attributed to a specific speaker may lead to severe social, legal, or economic consequences. Therefore, beyond generic detection systems, there is an urgent need for personalized, speaker-specific, and interpretable deepfake speech protection mechanisms.

\begin{figure}[t]
    \centering
\includegraphics[width=0.9\linewidth]{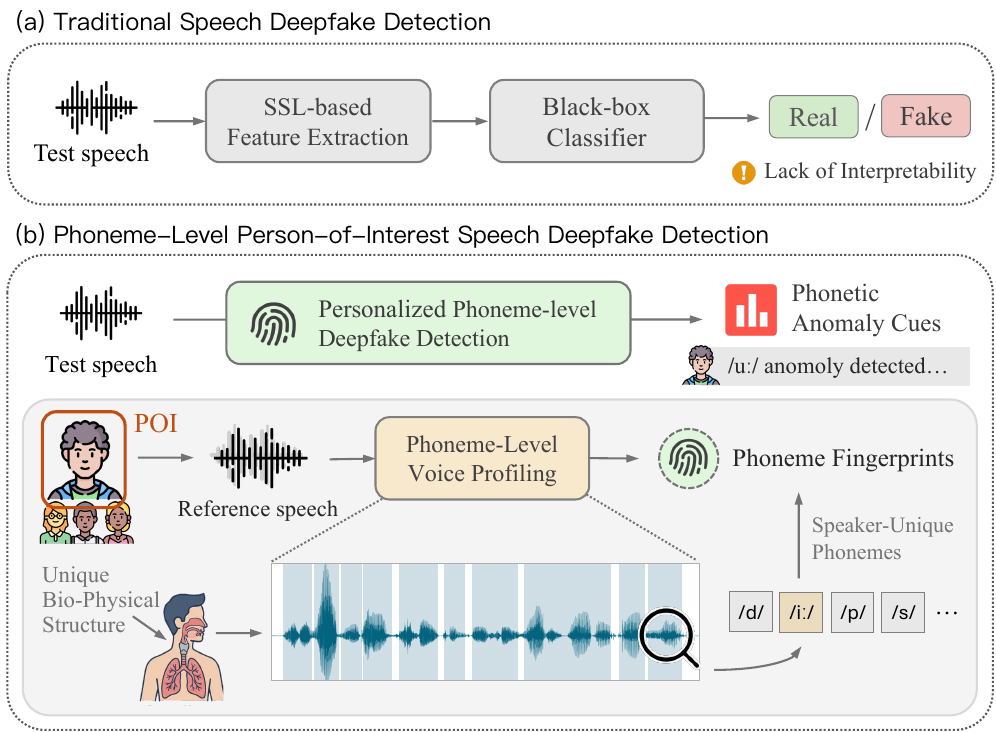}
    \caption{Illustration of our personalized and interpretable detection paradigm. (a) Traditional speech deepfake detection relies on black-box classifiers and lacks interpretability. (b) Our approach performs phoneme-level voice profiling based on the fact that each speaker's phoneme realizations are shaped by individual vocal characteristics and habitual articulation. By extracting speaker-unique phonemes and constructing ``Phoneme Fingerprints'' from reference speech, the system conducts granular verification on test speech and uncover ``Phonetic Anomaly Cues'', pinpointing specific suspicious phonetic segments for enhanced forensic transparency. }
    \label{fig:motivation}
\end{figure}

Most existing studies \cite{zhang2025multi,xue2026rtcfake,xue2023learning,xue2024dynamic} formulate speech deepfake detection as a straightforward classification problem, where an entire speech segment is processed by an end-to-end black-box model to output utterance-level decision. A common paradigm is to extract front-end representations using self-supervised learning (SSL) models, followed by a complex back-end classifier to perform discrimination (Fig~\ref{fig:motivation}a). Representative works~\cite{tak2022automatic,zhang2024audio,tran2025multi} adopt convolutional or transformer-based architectures to model global acoustic patterns and have achieved impressive performance on widely used benchmark datasets\cite{asvspoof,muller2022does}.

However, such evaluations are typically conducted under speaker-agnostic settings, where training and test speakers are loosely constrained. As a result, the learned decision boundaries tend to emphasize dataset-level artifacts—such as synthesis-model-specific spectral patterns, vocoder traces, or corpus-dependent biases—that are consistent across speakers but largely unrelated to individual articulation habits. In contrast, in POI scenarios, attackers often explicitly tailor the synthesis process to a single target speaker by fine-tuning generative models with bona fide speech from that individual \cite{famousfigures}. Under such conditions, generic detectors often struggle to capture subtle, speaker-dependent inconsistencies and provide limited interpretability regarding \emph{why} a given utterance is detected as fake.

To address this limitation, we turn our attention to the phoneme. Compared to frame-level artifacts that are often transient and model-dependent, phonemes offer a linguistically grounded unit whose acoustic realizations exhibit stable, speaker-specific patterns. Although phoneme categories are shared across speakers, their realizations are shaped by individual vocal characteristics and habitual articulation behaviors, leading to consistent inter-speaker differences. These fine-grained regularities are difficult for current speech synthesis and voice cloning systems to reproduce faithfully, often giving rise to systematic phoneme-level inconsistencies. While recent studies~\cite{zhang2025phoneme,baser25b_interspeech} have begun to exploit phoneme-related information to guide deepfake detection models, phonemes are still treated as auxiliary signals within speaker-agnostic, utterance-level classifiers. Such designs overlook the fact that phoneme realizations are inherently speaker-dependent and physiologically constrained, limiting their ability to support personalized analysis and interpretable detection in person-of-interest (POI) scenarios.

\begin{figure}[t]
    \centering
    \includegraphics[width=1.0\linewidth]{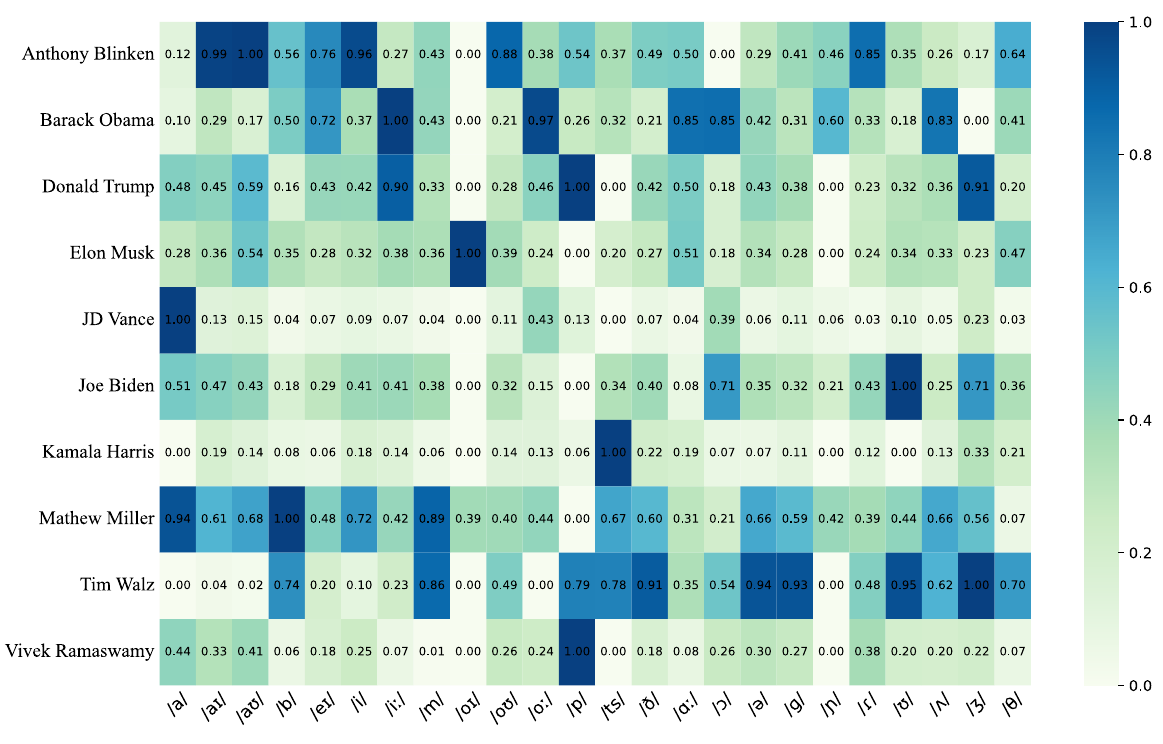} 
	\caption{Visualization of Speaker-Specific Phoneme Distinctiveness. Each cell shows $\mathrm{dist}(\mathbf{v}, \mathbf{c}) = 1 - \cos(\mathbf{v}, \mathbf{c})$, where $\mathbf{v}$ is the speaker's phoneme vector (averaged over frame-level embeddings) and $\mathbf{c}$ is the global centroid for that phoneme across all speakers. Darker cells indicate highly discriminative phonemes for the speaker, forming the fundamental motivation for our POI profiling. }

    \label{fig:heatmap}

\end{figure}

This intuition is further supported by a preliminary analysis on the \textit{FamousFigures} dataset. We sample genuine speech from ten speakers and extract phoneme-related features using the MMS model~\cite{ge2025post}, a multilingual speech deepfake detection system. These features are then visualized as heatmaps. As shown in Fig~\ref{fig:heatmap}, distinct phoneme pronunciation patterns can be observed across different speakers, indicating that phoneme-level representations capture stable and discriminative speaker-specific characteristics. These observations suggest that phonemes provide a natural and interpretable basis for personalized modeling in speech deepfake detection.

Building on these insights, we propose a novel, plug-and-play phoneme-level modeling framework for speaker-specific deepfake detection (Fig.~\ref{fig:motivation}b). Given a small amount of bona fide reference speech from a POI, our method first performs phoneme-level alignment and extracts phoneme-conditioned acoustic representations. For each phoneme, a lightweight Gaussian Mixture Model (GMM) is constructed to characterize the speaker’s habitual articulatory distribution. At inference time, a test utterance is decomposed into phonetic units and evaluated against the corresponding phoneme models, producing fine-grained consistency scores that reflect how well each phoneme realization matches the target speaker’s profile. These phoneme-level consistency scores expose localized speaker-dependent deviations, offering fine-grained and interpretable evidence for deepfake analysis.

Distinct from existing heavy-training paradigms, our method is agnostic to synthesis algorithms, thus reduces overfitting to spoof-specific artifacts, and generalizes naturally to unseen spoofing attacks. Moreover, its additive nature allows it to function as an interpretable ``plugin'' that can be seamlessly combined with existing detection backbones. Our contributions can be summarized as:

\begin{itemize}
\item \textbf{Speaker-Specific Phoneme Profiling:}
We propose a personalized deepfake detection framework that explicitly captures speaker-specific articulatory patterns using lightweight statistical modeling, enabling effective POI protection with minimal bona fide reference data and without relying on spoof-specific training. 

\item \textbf{Fine-Grained and Interpretable Phonetic Evidence:}
We introduce an interpretable phoneme-level scoring and tiered decision mechanism that reveals explicit phonetic anomaly cues, facilitating forensic analysis and transparent reasoning beyond black-box utterance-level classifiers.

\item \textbf{Chinese POI Deepfake Dataset:}
To fill the gap in speaker-centric benchmarks for Mandarin, we present the first large-scale Chinese POI deepfake dataset. It covers diverse public figures and modern speech generation techniques, enabling systematic evaluation of personalized spoofing attacks.
\end{itemize}

\section{Related Work}
\label{sec:related_work}

\subsection{Self-Supervised Learning for Speech Deepfake Detection}
The paradigm of speech deepfake detection has shifted significantly from handcrafted features ~\cite{xue2022audio,lavrentyeva2019stc} (e.g., LFCC, CQCC) to representation learning, driven by the success of Self-Supervised Learning (SSL) foundation models. Models such as wav2vec 2.0~\cite{baevski2020wav2vec}, HuBERT~\cite{hsu2021hubert}, and WavLM~\cite{chen2022wavlm} leverage massive unlabeled corpora to encode rich acoustic and phonetic information in a latent and entangled manner, proving highly effective at capturing subtle artifacts introduced by neural vocoders and upsampling layers. While Wav2Vec 2.0 utilizes contrastive predictive coding to retain phase information crucial for detecting synthesis discontinuities, WavLM explicitly incorporates a denoising objective, offering superior robustness against channel variation and background noise. Recent studies have further explored cross-lingual generalization using models like XLS-R~\cite{babu2021xlsr} to address the scarcity of non-English benchmarks, particularly for tonal languages like Mandarin. However, most SSL-based detectors utilize these embeddings as global, utterance-level representations, often discarding the fine-grained, localized phonetic discrepancies that are critical for identifying high-fidelity, targeted spoofing attacks.

\subsection{Speech Deepfake Detection: From Agnostic to Speaker-Specific}
Existing detection frameworks can be broadly categorized into speaker-agnostic and speaker-specific approaches. Speaker-agnostic methods aim to learn universal forensic features applicable to any voice. Common architectures in this domain, such as RawNet2~\cite{tak2021end} and AASIST~\cite{jung2022aasist}, employ sinc-convolutions and graph attention networks, respectively, to model spectro-temporal artifacts. Despite their success on standard benchmarks (e.g., ASVspoof), these ``black-box'' models often struggle with generalization against unseen generation algorithms and lack interpretability. 

Conversely, Speaker-Specific or POI detection reframes the task as a verification problem, leveraging reference data to protect specific identities. Early works~\cite{jung2022sasv} integrate automatic speaker verification with anti-spoofing to detect personalized clones, while recent advancements utilize speaker profiles to detect speaker-specific impersonation. Notably, emerging research has begun to investigate leveraging phoneme-level features for enhancing detection performance~\cite{zhang2025phoneme}, or performing phoneme-level analysis on POI~\cite{salvi2025phoneme}. Our work builds upon this interpretable direction by establishing a comprehensive Chinese POI benchmark and introducing an adaptive profiling mechanism that explicitly models these idiosyncratic articulatory patterns.

\section{Dataset Construction}
\label{sec:dataset}

To facilitate the study of POI deepfake detection under realistic conditions, we construct a large-scale speaker-centric Mandarin dataset containing both genuine and synthesized speech from public figures. The dataset is designed to benchmark personalized impersonation attacks, enabling fine-grained analysis of speaker-dependent artifacts.

\paragraph{Data Collection and Preprocessing.}
We collect approximately 400 hours of real speech from 10 target speakers on major Chinese online streaming platforms. To ensure acoustic consistency, all recordings are processed using a voice activity detection (VAD) tool to extract speech-only segments. Non-speech regions and excessively short or long segments are discarded.

To further guarantee speaker purity, we leverage pretrained automatic speaker verification (ASV) embeddings to verify speaker identity. Segments whose embeddings deviate significantly from the target speaker profile are removed, effectively filtering out background speakers, interviewers, and cross-talk commonly present in in-the-wild recordings.

\paragraph{Speech Synthesis.}
For each target speaker, we generate spoofed speech using five representative zero-shot text-to-speech (TTS) systems: F5-TTS, IndexTTS, LLaSA, OpenAudio-S1-mini, and VOXCPM. The selected TTS and VC systems are chosen to cover a wide range of modern zero-shot speech synthesis paradigms, including diffusion-based, autoregressive, and large language model (LLM) based architectures. All systems support reference-based voice cloning, enabling speaker-specific synthesis without fine-tuning. For each synthesis, a single genuine utterance from the target speaker is used as the reference signal, and the synthesis text is kept identical to the corresponding real speech. 

\paragraph{Post-processing.}
All real and synthesized audio samples are resampled to a unified sampling rate and undergo identical post-processing procedures to eliminate confounding factors introduced by format or codec differences. The resulting dataset provides a clean and balanced benchmark for evaluating speaker-specific deepfake detection methods.

\paragraph{Dataset Statistics and Composition.}
Table~\ref{tab:dataset-statistics} summarizes the overall statistics of the constructed dataset. The dataset covers ten target speakers with balanced real and synthesized utterances. For synthesized data, samples are evenly generated by multiple zero-shot TTS and VC systems to ensure diversity in generation mechanisms and artifacts.

\begin{table}[t]
\centering
\footnotesize
\setlength{\tabcolsep}{6pt}
\begin{tabular}{lcccc}
\toprule
\textbf{Item} & \textbf{Min} & \textbf{Max} & \textbf{Avg.} & \textbf{Total} \\
\midrule
Utt. per Speaker        & 1,848  & 71,080  & 20,494  & 204,944 \\
Real Utt. per Speaker  & 308   & 11,848  & 3,401  & 34,014 \\
Fake Utt. per Speaker  & 1,540  & 59,232  & 17,093  & 170,930 \\
Duration per Utt. (s)   & 1.25  & 50.00 & 7.25  & --     \\
Total Duration (h)     & --   & --   & --   & 412.48   \\
\bottomrule
\end{tabular}
\caption{Dataset statistics of the constructed speaker-specific spoofing detection dataset.}
\label{tab:dataset-statistics}
\end{table}

\begin{figure*}[ht]
    \centering
    \includegraphics[width=1\textwidth]{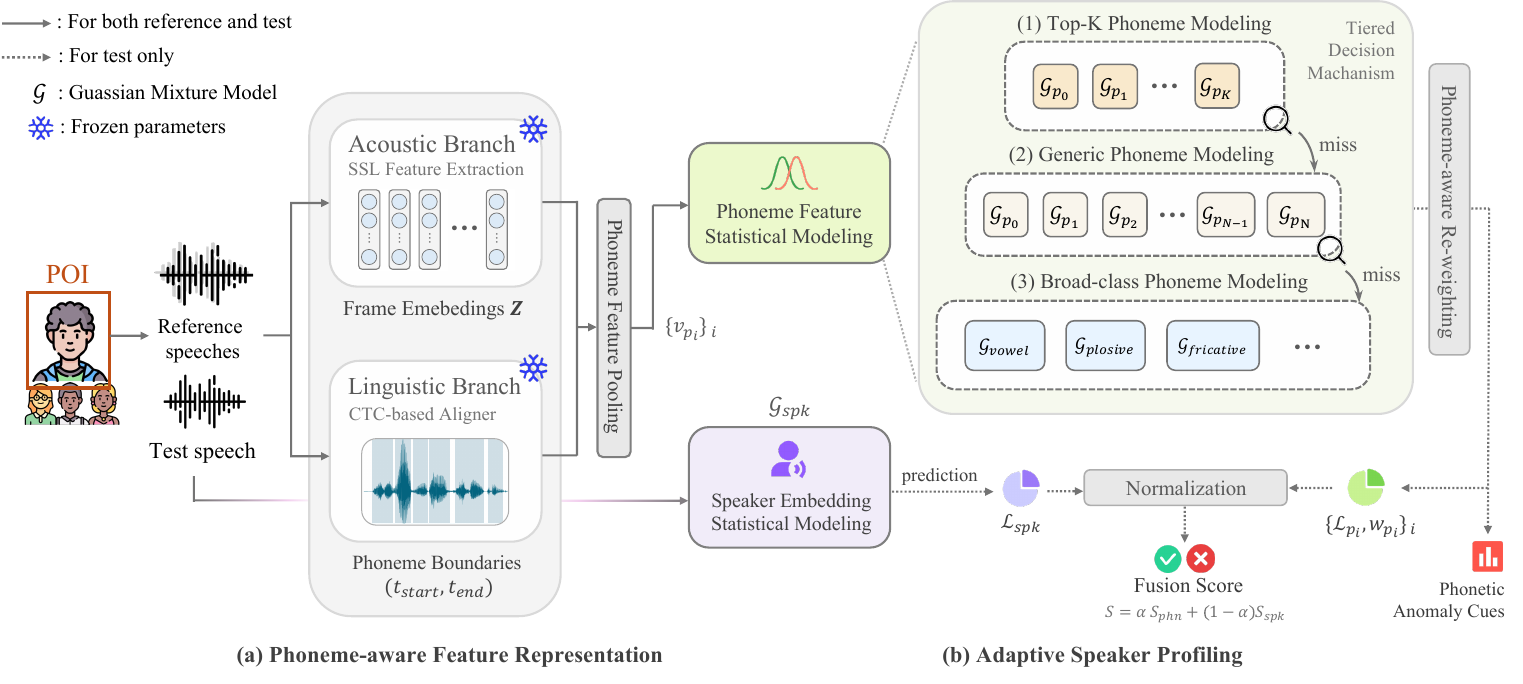} 
    \caption{The detailed architecture of our proposed personalized speech deepfake detection framework. The pipeline consists of two primary stages: (a) Phoneme-aware Feature Representation, which utilizes a dual-branch extractor to generate pooled phoneme vectors $\{v_{p_i}\}_i$; and (b) Adaptive Speaker Profiling, which models both fine-grained phonetic distributions and holistic speaker embeddings. A tiered decision mechanism (Top-K, Generic, and Broad-class) is employed to handle linguistic sparsity. Finally, a hybrid score $S$ is derived by fusing the phoneme-level likelihood $S_{phn}$ and the global identity score $S_{spk}$, enabling the identification of interpretable Phonetic Anomaly Cues.}
    \label{fig:framework}
\end{figure*}

\section{Methodology}
\label{sec:methodology}

We propose a speaker-specific and phoneme-aware speech deepfake detection framework designed for POI scenarios. Given a small amount of reference speech from a target speaker, our method constructs a personalized phoneme-level acoustic profile and performs interpretable detection by measuring phonetic consistency at inference time. 

\subsection{Phoneme-aware Feature Representation}
To achieve a comprehensive defense, we extract features at two distinct granularities: micro-level phoneme representations and macro-level speaker embeddings.

\textbf{Phoneme Feature Extraction:} We employ SSL model fine-tuned on speech deepfake datasets as the backbone $\mathcal{F}$. For a given input $x$, the model extracts a sequence of high-dimensional acoustic embeddings:
\begin{equation}
    \mathbf{Z} = \mathcal{F}(x) = [\mathbf{z}_1, \mathbf{z}_2, \dots, \mathbf{z}_T], \quad \mathbf{z}_t \in \mathbb{R}^D
\end{equation}
where $T$ is the number of frames and $D$ is the embedding dimension. These features encompass rich prosodic and phonetic information learned from diverse linguistic contexts.

Simultaneously, a Connectionist Temporal Classification (CTC) based aligner $\mathcal{A}$ is utilized to determine the phonetic boundaries. For each phoneme $p$ appearing in the utterance, the aligner provides its temporal interval $[t_{start}, t_{end}]$. To obtain a fixed-length representation for each phoneme instance, we apply a mean-pooling operation over the corresponding frame indices:
\begin{equation}
    \mathbf{v}_p = \frac{1}{t_{end} - t_{start} + 1} \sum_{t=t_{start}}^{t_{end}} \mathbf{z}_t
\end{equation}
The resulting vector $\mathbf{v}_p$ serves as a micro-level descriptor of the speaker's idiosyncratic pronunciation for phoneme $p$. 

\textbf{Global Speaker Embedding:} To capture the holistic vocal identity, we utilize a pre-trained ASV model, denoted as $\mathcal{F}_{spk}$. For the input utterance $x$, we extract a fixed-dimensional global speaker embedding:
\begin{equation}
    \mathbf{e} = \mathcal{F}_{spk}(x), \quad \mathbf{e} \in \mathbb{R}^{D_{spk}}
\end{equation}
This embedding encapsulates long-term spectral and prosodic characteristics, providing a robust identity anchor complementary to the transient phonetic features.

\subsection{Adaptive Speaker Profiling}
\label{sec:profiling}

A single phoneme may exhibit multiple articulation modes for the same speaker due to coarticulation, prosodic variation, and speaking style. To explicitly model such intra-speaker phonetic variability, we employ GMMs to model the distribution of each phoneme in the speaker’s acoustic space. This formulation enables flexible modeling of intra-speaker variability while retaining a compact and interpretable statistical representation.

\textbf{Phoneme Statistical Modeling:} Let $\mathcal{E}_S = \{ x_i \}_{i=1}^{N}$ denote the reference utterances of the target speaker. For each unique phoneme $p$ in the reference set $\mathcal{E}_S$, we fit a GMM with $K_p$ components. The probability density function for a phoneme vector $\mathbf{v}$ is defined as:
\begin{equation}
    P(\mathbf{v} | \lambda_p) = \sum_{k=1}^{K_p} \pi_k \mathcal{N}(\mathbf{v} | \boldsymbol{\mu}_k, \boldsymbol{\Sigma}_k)
\end{equation}
where $\lambda_p = \{ \pi_k, \boldsymbol{\mu}_k, \boldsymbol{\Sigma}_k \}_{k=1}^{K_p}$ denotes the model parameters, including mixture weights $\pi_k$, mean vectors $\boldsymbol{\mu}_k$, and covariance matrices $\boldsymbol{\Sigma}_k$. To prevent over-fitting on sparse data, we restrict $\boldsymbol{\Sigma}_k$ to be a diagonal covariance matrix and adaptively adjust $K$ based on the sample size $N_p$ of the phoneme.

\textbf{Profile Reliability Weighting:} Not all phonemes are equally discriminative or stable for a given speaker. To quantify the reliability of a phoneme profile, we compute a confidence weight $w_p$ based on the average log-likelihood ($\bar{\mathcal{L}}_p$) of the reference samples:
\begin{equation}
    \bar{\mathcal{L}}_p = \frac{1}{N_p} \sum_{i=1}^{N_p} \log P(\mathbf{v}_i | \lambda_p)
\end{equation}
The weight $w_p$ is then derived using an exponential scaling factor $\alpha$:
\begin{equation}
    w_p = \exp(\bar{\mathcal{L}}_p / \alpha)
\end{equation}
A higher $w_p$ indicates that the speaker’s pronunciation of phoneme $p$ is highly consistent (compact in the feature space), making it a more reliable "fingerprint" for subsequent deepfake detection.

\textbf{Global Identity Modeling:} To model the speaker's holistic variance, we fit a separate GMM, denoted as $\lambda_{spk}$, to the set of global embeddings $\{\mathbf{e}_i\}_{i=1}^N$ extracted from the reference utterances $\mathcal{E}_S$. Similar to the phonetic branch, we employ a diagonal covariance matrix to prevent overfitting on limited data:
\begin{equation}
    P(\mathbf{e} | \lambda_{spk}) = \sum_{k=1}^{K_{spk}} \pi_k \mathcal{N}(\mathbf{e} | \boldsymbol{\mu}_k, \boldsymbol{\Sigma}_k)
\end{equation}
where $K_{spk}$ is number of mixture components. This global model serves as a coarse-grained verifier to ensure the test utterance falls within the target speaker's general acoustic manifold.

\subsection{Identification of Salient Phoneme Fingerprints}

Phonemes differ in their stability and speaker-discriminative capacity. Rather than treating all phonetic units equally, we identify a speaker-specific subset of \emph{salient phoneme fingerprints}, denoted as $\mathcal{P}_{salient}$, which capture the most consistent articulatory patterns of the POI.

Let $\mathcal{P}_{all}$ denote the set of phoneme types observed in the reference utterances of the target speaker. As described in Algorithm~\ref{alg:alg1}, the selection of salient phonemes is driven by the phoneme reliability weights $w_p$ introduced in Section~\ref{sec:profiling}. Specifically, phonemes in $\mathcal{P}_{all}$ are ranked according to $w_p$, and only the top-$K$ most reliable units are retained to form $\mathcal{P}_{salient}$. By focusing on phonemes that exhibit high intra-speaker consistency, this selection yields a compact, speaker-specific, and interpretable phonetic fingerprint.

\begin{algorithm}[ht]
\caption{Speaker-specific Salient Phoneme Modeling and Tiered Decision}
\label{alg:alg1}
\begin{algorithmic}[1]
    \renewcommand{\algorithmicrequire}{\textbf{Input:}}
    \renewcommand{\algorithmicensure}{\textbf{Output:}}
    
    \REQUIRE Enrollment set $\mathcal{P}_{\text{all}}$, weights $\{w_p\}$, test utterance $X$, detected phonemes $\mathcal{P}_{\text{test}}$, parameter $K$.
    \ENSURE Detection score $S(X)$.

    \vspace{0.4em}
    \item[\color{gray}\textit{\quad// Salient Phoneme Fingerprint Construction}]
    
    \STATE $\mathcal{P}_{\text{sorted}} \gets \text{Sort}(\mathcal{P}_{\text{all}}, \{w_p\}, \text{desc})$ 
    \STATE $\mathcal{P}_{\text{salient}} \gets \text{SelectTopK}(\mathcal{P}_{\text{sorted}}, K)$ 

    \vspace{0.4em}
    \item[\color{gray}\textit{\quad// Tiered Decision Mechanism}]
    
    \IF{$\mathcal{P}_{\text{test}} \cap \mathcal{P}_{\text{salient}} \neq \emptyset$}
        \STATE $S(X) \gets \text{WeightedAvg}\left( \{s_p\}_{p \in \mathcal{P}_{\text{test}} \cap \mathcal{P}_{\text{salient}}}, \{w_p\} \right)$
        
    \ELSIF{$\mathcal{P}_{\text{test}} \cap \mathcal{P}_{\text{all}} \neq \emptyset$}
        \STATE $S(X) \gets \text{Avg}\left( \{s_p\}_{p \in \mathcal{P}_{\text{test}} \cap \mathcal{P}_{\text{all}}} \right)$
        
    \ELSE
        \STATE $\mathcal{C}_{\text{test}} \gets \{ C(p) \mid p \in \mathcal{P}_{\text{test}} \}$
        \STATE $S(X) \gets \text{Avg}\left( \{ s_c \}_{c \in \mathcal{C}_{\text{test}}} \right)$
    \ENDIF

    \vspace{0.4em}
    \RETURN $S(X)$
\end{algorithmic}
\end{algorithm}

\subsection{Tiered Decision Mechanism}

In practical POI scenarios, test utterances are often short or linguistically constrained, leading to incomplete phoneme coverage. Let $\mathcal{P}_{\text{test}}$ denote the set of phonemes detected in a test utterance. Relying solely on a fixed set of salient phonemes may therefore result in unreliable or missing decisions. To address this issue, we adopt a tiered decision mechanism that adaptively selects the most informative level of phonetic evidence available at inference time.

As illustrated in Algorithm~\ref{alg:alg1}, the decision process proceeds in a coarse-to-fine manner. If $\mathcal{P}_{\text{test}}$ contains any speaker-specific salient phonemes, the final score is computed as a reliability-weighted average of their normalized phoneme-level scores $s_p$, yielding a high-precision and interpretable decision. When no salient phonemes are observed but other reference phonemes are present, the system falls back to a weighted average over all available phoneme-level scores. In the extreme case where phoneme overlap is sparse, phonemes are mapped to broader phonetic categories via a predefined function $C(\cdot)$ (e.g., vowels, plosives or fricatives), and category-level scores $s_c$ are used to ensure coverage.

This tiered approach prioritizes discriminative, speaker-consistent phonemes whenever possible, while maintaining robustness against linguistic sparsity and content mismatch.

\subsection{Hybrid Scoring and Fusion Strategy}
Before combining phoneme-level evidence, we normalize raw likelihood scores to ensure comparability across phonemes. For each phoneme $p$, we compute its log-likelihood $\mathcal{L}_p = \log P(\mathbf{v}_p \mid \lambda_p)$ under the corresponding phoneme GMM, and apply a Sigmoid-based normalization:
\begin{equation}
    s_p = \sigma(\mathcal{L}_p; \beta_{spk}, \gamma_{spk}) = \frac{1}{1 + \exp(-(\mathcal{L}_p - \beta_{spk}) / \gamma_{spk})}
\end{equation}
where $\beta$ and $\gamma$ are the centering and scaling hyper-parameters determined empirically.

Finally, we fuse phoneme-level evidence with global speaker identity to produce the final detection score. We employ the tiered decision mechanism to compute the phoneme-level score $S_{phn}(X)$.  Parallelly, we compute the log-likelihood of the test utterance's global embedding $\mathbf{e}_{test}$ given the global profile $\lambda_{spk}$, followed by Sigmoid normalization:
\begin{equation}
    S_{spk}(X) = \sigma(\log P(\mathbf{e}_{test} \mid \lambda_{spk}); \beta_{spk}, \gamma_{spk})
\end{equation}
where $\beta_{spk}$ and $\gamma_{spk}$ are normalization parameters centered on the impostor distribution.

To balance the sensitivity to local artifacts and the robustness of global identity, the final detection score $S_{final}$ is obtained via linear interpolation:
\begin{equation}
    S_{final}(X) = \alpha \cdot S_{phn}(X) + (1 - \alpha) \cdot S_{spk}(X)
\end{equation}
where $\alpha \in [0, 1]$ controls the trade-off between phoneme-level and speaker-level evidence.

\begin{table*}[t]
    \centering
    \label{tab:backbone_vp_zh_en_pooled}
    \setlength{\tabcolsep}{4pt}
    \renewcommand{\arraystretch}{1.05}
    \begin{tabular}{c cc cc cc cc cc cc}
        \specialrule{1.2pt}{0pt}{3pt}
        \multirow{4}{*}{\textbf{Backbone}}
        & \multicolumn{4}{c}{\textbf{ZH-Famous}}
        & \multicolumn{4}{c}{\textbf{En-Famous}}
        & \multicolumn{4}{c}{\textbf{Pooled}} \\
        
        \cmidrule(lr){2-5}\cmidrule(lr){6-9}\cmidrule(lr){10-13}
        & \multicolumn{2}{c}{Baseline} & \multicolumn{2}{c}{with PVP}
        & \multicolumn{2}{c}{Baseline} & \multicolumn{2}{c}{with PVP}
        & \multicolumn{2}{c}{Baseline} & \multicolumn{2}{c}{with PVP} \\
        
        \cmidrule(lr){2-3}\cmidrule(lr){4-5}
        \cmidrule(lr){6-7}\cmidrule(lr){8-9}
        \cmidrule(lr){10-11}\cmidrule(lr){12-13}
        & {\scriptsize AUC ($\uparrow$)} & {\scriptsize EER ($\downarrow$)}
        & {\scriptsize AUC ($\uparrow$)} & {\scriptsize EER ($\downarrow$)}
        & {\scriptsize AUC ($\uparrow$)} & {\scriptsize EER ($\downarrow$)}
        & {\scriptsize AUC ($\uparrow$)} & {\scriptsize EER ($\downarrow$)}
        & {\scriptsize AUC ($\uparrow$)} & {\scriptsize EER ($\downarrow$)}
        & {\scriptsize AUC ($\uparrow$)} & {\scriptsize EER ($\downarrow$)} \\
        \midrule
        
        hubert-xlarge
        & 63.70 & 39.80
        & 89.36 & 19.21
        & 86.88 & 19.76
        & 93.81 & 11.96
        & 75.29 & 29.78
        & 91.58 & 15.58 \\
				
		wav2vec2-small
        & 74.39 & 33.08
        & 88.84 & 19.37
        & 84.58 & 24.16
        & \cellcolor{secondgray}\underline{96.17} & 9.93
        & 79.48 & 28.62
        & 92.50 & 14.65 \\
		
		wav2vec2-large
        & 78.81 & 28.19
        & 91.81 & 15.95
        & 93.93 & 12.66
        & \cellcolor{bestgray}\textbf{98.22} & \cellcolor{bestgray}\textbf{4.46}
        & 86.37 & 20.42
        & 95.01 & 10.20 \\
		
		wav2vec2-xlsr-1b
        & 88.43 & 19.17
        & \cellcolor{secondgray}\underline{92.93} & \cellcolor{secondgray}\underline{14.91}
        & 95.65 & 11.30
        & 98.21 & 6.66
        & 92.04 & 15.23
        & \cellcolor{secondgray}\underline{95.57} & \cellcolor{secondgray}\underline{10.78} \\

        mms-300m
        & 84.11 & 21.13
        & \cellcolor{bestgray}\textbf{94.58} & \cellcolor{bestgray}\textbf{11.37}
        & 93.99 & 13.97
        & 96.61 & \cellcolor{secondgray}\underline{7.24}
        & 89.05 & 17.55
        & \cellcolor{bestgray}\textbf{95.59} & \cellcolor{bestgray}\textbf{9.305} \\
        
        \specialrule{1.2pt}{0pt}{0pt}
    \end{tabular}
    \caption{Plug-and-play evaluation of the proposed Phoneme-based Voice Profiling (PVP) module with different backbone feature extractors on two POI datasets, reported in terms of AUC (↑) / EER (↓) (\%).}
\label{tab:full-comparison}
\end{table*}

\section{Experiments} \label{sec:experiments}

\subsection{Implementation Details}

\textbf{Datasets.}
We evaluate our method on two POI deepfake datasets to assess its effectiveness and robustness under diverse linguistic and attack conditions. In addition to our proposed Chinese POI dataset, we adopt the \textit{Famous Figures} dataset~\cite{famousfigures}, an English POI deepfake benchmark featuring diverse speech synthesis attacks under zero-shot, few-shot, and fine-tuned settings. Both our dataset (denoted as \textbf{ZH-Famous}) and \textit{Famous Figures} (denoted as \textbf{EN-Famous}) are designed for POI scenario and include previously unseen synthesis algorithms, making them well suited for assessing speaker-centric and attack-agnostic defenses.

\textbf{Model Configuration.}
For phoneme-level alignment, we employ a \texttt{wav2vec2-large-xlsr-53} model\footnote{\url{https://huggingface.co/facebook/wav2vec2-xlsr-53-espeak-cv-ft}} to extract phoneme boundary timesteps. For speaker-level representation, we adopt a pretrained \texttt{ECAPA-TDNN} model\footnote{\url{https://huggingface.co/speechbrain/spkrec-ecapa-voxceleb}} to extract fixed-dimensional global speaker embeddings.

Each phoneme is modeled using a GMM with diagonal covariance matrices. Unless otherwise specified, we utilize 1\% of the bona fide speech from each speaker's available data as reference data for all experiments. We fix the number of mixture components to $K_p=5$ for phoneme-level models and $K_{spk}=5$ for the global speaker model, together with a covariance regularization term of $10^{-3}$ for numerical stability. The number of salient phonemes is set to $K=12$ across all experiments. For likelihood normalization, we apply a sigmoid mapping with $\beta=-2000$ and $\gamma=200$. For final score fusion, we set $\alpha=0.8$.

\textbf{Evaluation Metrics.}
We report Equal Error Rate (EER, \%) and Area Under the ROC Curve (AUC, \%) as the primary evaluation metrics. Lower EER and higher AUC indicate better detection performance.

\subsection{Plug-and-Play Evaluation}
A core design goal of the proposed Phoneme-based Voice Profiling (PVP) module is plug-and-play compatibility with existing deepfake detection backbones. To validate this property, we integrate PVP with a diverse set of SSL-based encoders~\cite{ge2025post} and evaluate performance on POI and cross-lingual benchmarks.


As summarized in Table~\ref{tab:full-comparison}, incorporating PVP consistently improves detection performance across all backbones on both datasets. On the EN-Famous benchmark, PVP integration yields an average EER reduction of approximately $8.3\%$ and an average AUC improvement of about $5.6\%$ across different backbones. On the ZH-Famous dataset, PVP brings more substantial gains, achieving an average EER reduction of around $12.1\%$ and an average AUC increase of approximately $13.6\%$, demonstrating strong effectiveness under cross-lingual and unseen generation conditions.

\subsection{Comparison with State-of-the-Art Methods}

We compare our method with state-of-the-art deepfake detection approaches under a unified speaker-centric evaluation protocol. As shown in Table~\ref{tab:sota-comparison}, the proposed method consistently achieves the best performance across both benchmarks, with substantially higher AUC and lower EER than all competing approaches.

\begin{table}[t]
	\centering
    \footnotesize

	\setlength{\tabcolsep}{1pt}
	\renewcommand{\arraystretch}{1.05}
	\begin{tabular}{l cc cc}
		\specialrule{1.2pt}{0pt}{3pt}
        \multirow{2}{*}[-4pt]{\textbf{Model}}

        & \multicolumn{2}{c}{\small\textbf{ZH-Famous}} 
        & \multicolumn{2}{c}{\small\textbf{EN-Famous}} \\
        \cmidrule(lr){2-3}\cmidrule(lr){4-5}
        & {\scriptsize AUC ($\uparrow$)} & {\scriptsize EER ($\downarrow$)} 
        & {\scriptsize AUC ($\uparrow$)} & {\scriptsize EER ($\downarrow$)} \\
        \midrule

		LCNN~\cite{lcnn}            & 41.73 & 53.83 & 43.91 & 53.81 \\
		RawNet2~\cite{tak2021end}         & 46.51 & 50.84 & 35.09 & 59.01 \\
		RawGAT-ST~\cite{rawgatst}       & 38.17 & 58.94 & 27.67 & 68.67 \\
		LibriSeVoc~\cite{librisevoc}      & 55.12 & 46.85 & 50.48 & 50.40 \\
		
		AASIST~\cite{jung2022aasist}          & 33.27 & 60.90 & 34.04 & 62.83 \\
		XLSR+AASIST~\cite{tak2022automatic}     & 44.54 & 50.95 & 45.72 & 53.06 \\
		XLSR+SLS~\cite{zhang2024audio}        & 54.55    & 42.65    & 50.14 & 45.62 \\
		
		ML-SSLFG~\cite{tran2025multi}       
		& 45.24 
		& 50.41 
		& \cellcolor{secondgray}\underline{64.49} 
		& \cellcolor{secondgray}\underline{39.50} \\

		PLFD-ADD~\cite{zhang2025phoneme} 
		& \cellcolor{secondgray}\underline{61.63} 
		& \cellcolor{secondgray}\underline{41.74} 
		& 47.42 & 51.85 \\
		
		
		\rowcolor{bestgray}
		\textbf{PVP (Ours)}   
		& \textbf{94.58} 
		& \textbf{11.37} 
		& \textbf{96.61} 
		& \textbf{7.24} \\
		
		\specialrule{1.2pt}{0pt}{0pt}
	\end{tabular}
    \caption{Comparison with state-of-the-art methods on POI datasets. Results are reported as AUC (↑) / EER (↓) (\%).}
    \label{tab:sota-comparison}
\end{table}

Compared to general-purpose detectors, our framework exhibits a clear advantage in capturing fine-grained, speaker-dependent articulatory patterns that are critical for detecting POI speech synthesis. In particular, the large performance margins indicate that modeling phoneme-level consistency from bona fide speech provides strong robustness against unseen spoofing attacks, even without exposure to attack-specific data during modeling. Despite being designed for personalized defense, the method also maintains strong performance across both English and Chinese benchmarks, demonstrating favorable cross-lingual generalization.

\subsection{Ablation Study}
\label{subsec:ablation}

We conduct ablation studies to examine the impact of key design choices in the proposed PVP framework under the same POI evaluation protocol. Specifically, we replace phoneme-level GMM likelihood modeling with cosine similarity between phoneme embeddings, remove the global speaker embedding branch, and discard phoneme-level profiling in favor of utterance-level modeling, respectively.

\begin{table}[t]
    \centering
    \footnotesize
    \setlength{\tabcolsep}{4pt}
    \renewcommand{\arraystretch}{1.0}
    \begin{tabular}{l cc cc}
    \toprule
    \multirow{2}{*}[-2pt]{\textbf{Method Variant}} 
    & \multicolumn{2}{c}{\textbf{ZH-Famous}} 
    & \multicolumn{2}{c}{\textbf{EN-Famous}} \\
    \cmidrule(lr){2-3} \cmidrule(lr){4-5}
    & {\scriptsize AUC ($\uparrow$)} & {\scriptsize EER ($\downarrow$)}
    & {\scriptsize AUC ($\uparrow$)} & {\scriptsize EER ($\downarrow$)} \\
    \midrule
    w/o Phoneme Modeling        & 92.80 & 14.95 & 94.98 & 9.62 \\
    w/o GMM        & 92.70 & 14.49 & 98.16 & 7.20 \\
    w/o Speaker Embedding           & 92.78 & 13.44 & 98.42 & 7.43 \\
    \midrule
    \textbf{Full PVP (Ours)}        & 94.58 & 11.37 & 96.61 & 7.24 \\
    \bottomrule
    \end{tabular}
    \caption{Ablation study on the POI benchmarks. Results are reported as AUC (↑) / EER (↓) (\%).}
    \label{tab:ablation}
\end{table}

Table~\ref{tab:ablation} shows that each component of PVP contributes positively to detection performance under the POI benchmarks. Removing phoneme-level profiling and reverting to utterance-level modeling causes the most consistent degradation, confirming the importance of fine-grained phonetic analysis for capturing speaker-specific articulatory regularities. While certain components may exhibit slightly better performance under specific conditions, their effectiveness is not consistently preserved across datasets and metrics. In contrast, the full PVP framework, which integrates phoneme-level modeling, probabilistic characterization, and global speaker cues, delivers the most stable and averaged-best performance. These results suggest that the strength of PVP lies not in any single module, but in the complementary interplay among its components.

\subsection{Interpretability and Case Study}

Beyond detection accuracy, our proposed PVP framework provides inherent interpretability by explicitly decomposing the final decision into phoneme-level, speaker-conditioned evidence. Unlike conventional end-to-end detectors that output utterance-level black-box scores, PVP allows each phonetic unit to be independently evaluated against the POI’s articulatory profile learned from bona fide reference speech, making the detection process transparent and traceable.

\begin{figure}[t]
    \centering
    \includegraphics[width=1\linewidth]{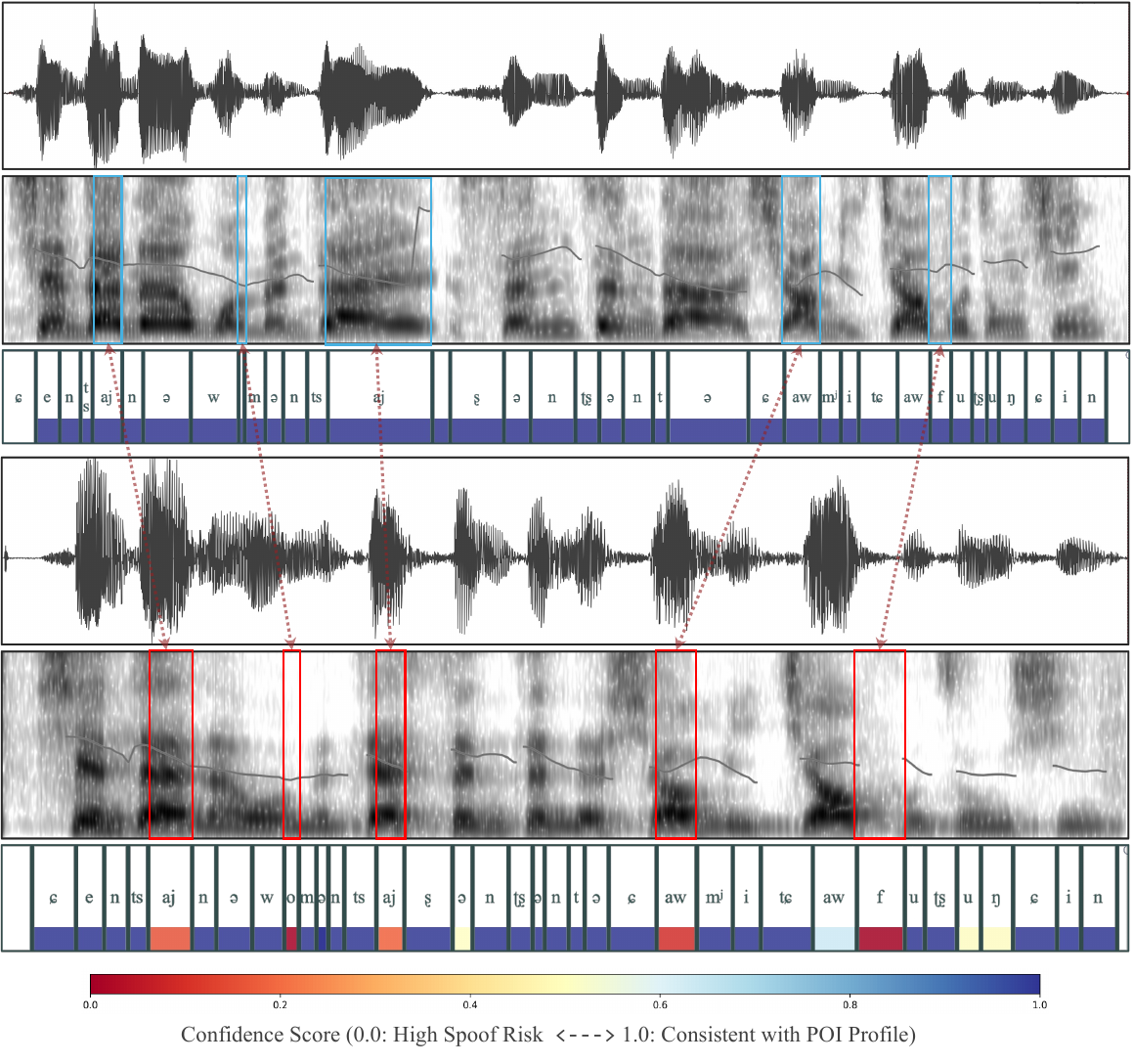} 
	\caption{Visualization of Phonetic Interpretability and Anomaly Detection. This figure compares a bona fide utterance (top) with a synthetic clone (bottom) of the same linguistic content. The segments are time-aligned with their respective waveforms, spectrograms, and phoneme labels. The confidence heatmap at the bottom of each panel reflects the consistency of each phoneme with the POI's articulatory profile: Blue (1.0) indicates high consistency, while Red (0.0) denotes a high spoofing risk.}
    \label{fig:case_study}

\end{figure}

This focus on phoneme-level speaker inconsistency provides actionable forensic cues while preserving robustness and generalization to previously unseen attacks. Such interpretability arises naturally from the design of PVP rather than post-hoc analysis: by enforcing phoneme-level alignment and probabilistic modeling, detection decisions are grounded in deviations from a speaker’s own phonetic distribution instead of attack-specific patterns. As illustrated in Fig~\ref{fig:case_study}, bona fide speech maintains consistently high confidence across segments, whereas the synthetic clone exhibits localized ``red flags'' at specific phonemes where the generative model fails to replicate the POI’s unique articulatory habits. Moreover, such structured, phoneme-level evidence offers a promising interface for future reasoning-based or large-model-driven forensic systems, where explicit articulatory consistency can serve as interpretable intermediate signals.

\section{Conclusion} \label{sec:conclusion}

This paper introduces Phoneme-based Voice Profiling (PVP), a plug-and-play and lightweight framework for POI speech deepfake detection. By shifting detection from black-box utterance-level classification to phonetic consistency verification, PVP models speaker-specific articulatory patterns as personalized ``fingerprints'' using lightweight GMM-based statistics. This design enables high data efficiency and strong generalization to unseen synthesis algorithms, without relying on heavy spoof-specific training.

Extensive evaluations on both our newly curated ZH-Famous dataset and \textit{Famous Figures} dataset demonstrate that PVP consistently outperforms state-of-the-art methods across diverse SSL backbones, achieving substantial reductions in EER. Beyond detection performance, the proposed framework provides transparent phonetic anomaly cues by exposing speaker-centric inconsistencies at the phoneme level, thereby bridging automated deepfake detection with interpretable POI forensic analysis. These findings suggest that phoneme-level voice profiling offers a principled and extensible direction for personalized and interpretable defense against POI speech synthesis.

\section*{Acknowledgements}
This work is supported by the Natural Science Foundation of China (NSFC) under the grant NO.62572358, 62372334

\nocite{ecapa}
\nocite{gmm}
\bibliographystyle{named}
\bibliography{ijcai26}

\end{document}